# Dimensional Control of Excitonic Interactions in Exfoliated 2D Molecular Crystals

Jonghyun Son[1], Seonghyun Koo[1], Daniel Yim[2,3], Sangjin Han[1], Dong-Hwan Yang[4], Gi-Yeop Kim[4], Kihyun Lee[5], Jieun Yeon[5], Hye Soo Kim[1], Eunbeen Jeon[1], Minji Ko[1], Minhee Choe[1], Kenji Watanabe[6], Takashi Taniguchi[7], Hee Cheul Choi[1], Kwanpyo Kim[5], Si-Young Choi[4], Seogjoo J. Jang[8], Hyungjun Kim[2,3] and Sunmin Ryu[1]*

*E-mail: sunryu@postech.ac.kr

[1]Department of Chemistry, Pohang University of Science and Technology (POSTECH), Pohang 37673, Republic of Korea

[2]Department of Chemistry, Hanyang University, Seoul 04763, Republic of Korea

[3]Research Institute of Natural Science, Hanyang University, Seoul 04763, Republic of Korea

[4]Department of Materials Science and Engineering, Pohang University of Science and Technology (POSTECH), Pohang 37673, Republic of Korea

[5]Department of Physics, Yonsei University, Seoul 03722, Republic of Korea

[6]Research Center for Materials Nanoarchitectonics, National Institute for Materials Science, 1-1 Namiki, Tsukuba 305-0044, Japan

[7]Research Center for Electronic and Optical Materials, National Institute for Materials Science, 1-1 Namiki, Tsukuba 305-0044, Japan

[8]Department of Chemistry and Biochemistry, Queens College, City University of New York, Queens, NY 11367, USA; PhD Programs in Chemistry and Physics, Graduate Center, City University of New York, New York, NY 10016, USA

## ABSTRACT

Excited states in molecular crystals depend on intermolecular coupling, charge-transfer interactions, dielectric screening and molecular packing. How reduced dimensionality modifies these interactions when molecular packing structure is preserved remains unclear. Here we demonstrate that tetracene and three other molecular crystals can be mechanically exfoliated into mono-, few- or multilayer flakes, exploiting the top-down approach established for inorganic layered materials. Electron diffraction confirms that exfoliated tetracene retains long-range crystalline order and nearly bulk-like in-plane lattice parameters. This structural fidelity enables systematic thickness-dependent spectroscopy with limited changes in molecular packing. With decreasing thickness, the Davydov splitting diminishes, whereas the Stokes shift increases to nearly three times its bulk-like value, indicating enhanced excited-state relaxation. The photoluminescence 0-0/0-1 intensity ratio also rises, suggesting greater exciton delocalization. Compared with exfoliated tetracene, assembly-grown crystals exhibit lattice contraction and larger Davydov splitting. Analysis within the Frenkel-charge-transfer mixing framework qualitatively accounts for the Davydov-state energy shifts and splitting in terms of molecular packing, intermolecular coupling and dielectric screening. Our results show that intermolecular interactions, excited-state relaxation and exciton coherence can be systematically tuned by varying layer number, demonstrating the feasibility and utility of extending dimensional control of inorganic 2D materials to molecular crystals.

## Introduction

Optical excitation of molecular crystals produces molecular excitons[1, 2] with Frenkel character, whose properties depend on non-covalent intermolecular interactions[3]. Consequently, optical and excited-state properties are highly sensitive to molecular packing[4], orbital overlap[5], polarization screening[6] and thermal lattice fluctuations[7]. These interactions influence characteristic phenomena including Davydov splitting (DS)[8, 9], superradiance[10], vibronic progression[11], Frenkel-charge transfer (CT) mixing[3, 9] and singlet fission[12], which are relevant to light harvesting, emissive devices, exciton transport and emerging photonic technologies. Because interactions in these molecular crystals are intrinsically weak and dependent on geometry and dielectric screening, reducing dimensionality can provide an effective route to tune molecular excitons with layer-by-layer precision. In the few-layer limit, reduced dielectric screening[13], modified lattice order[14] and broken out-of-plane translational periodicity[15] can perturb the balance among these intermolecular interactions and thereby reshape molecular excited states. Although analogous dimensional effects have been widely explored in inorganic 2D semiconductors and van der Waals (vdW) heterostructures[16-18], their counterparts for molecular crystals remain largely unexplored.

We ask whether reducing layer number can modify collective excited states in organic molecular crystals without substantially changing the underlying molecular packing[4, 19, 20]. Addressing this question requires molecular-layer thickness control with independent structural characterization. Most previous investigations have relied on multiple quantum wells[19], evaporated thin films[7] or epitaxially grown layers[21], where polymorphism[8], structural disorder[22], substrate-specific growth[23] or limited layer-number precision[19] can obscure intrinsic dimensional effects. Recently, Zhang et al. used argon-plasma thinning to examine thickness-dependent emission in pentacene, attributing the spectral changes to a transition between Frenkel and CT excitons[24]. Mechanical exfoliation[16], which transformed the study of inorganic layered materials, offers an attractive alternative because it can generate clean crystals with well-defined thicknesses and minimal chemical perturbation. Despite a few reports[25-27], however, whether molecular solids held together only by relatively weak van der Waals interactions can be exfoliated down to the monolayer or few-layer limit while preserving crystalline order remains uncertain.

Among molecular crystals, tetracene (Tc) serves as an ideal case study because its singlet excitons exhibit pronounced optical anisotropy[28], Davydov-split states[29], strong vibronic coupling[5] and intriguing excited-state dynamics[30]. These excitons are governed by intermolecular Coulomb interactions and mixing with charge-transfer states,[3] and should therefore respond sensitively to reduced dimensionality and the accompanying changes in environmental polarization or dielectric screening. Here, we show that Tc molecular crystals can be mechanically exfoliated into highly crystalline flakes down to the bilayer limit and pentacene (Pc) to the monolayer limit. Similar exfoliation is confirmed for diindenoperylene (DIP) and phenothiazine (PTZ), demonstrating the applicability of mechanical exfoliation beyond Tc. Electron diffraction and polarization-resolved optical analyses reveal that long-range molecular order is preserved after exfoliation. Thickness-dependent spectroscopy demonstrates that layer-number reduction continuously modulates the characteristics of molecular excitons, which reflect changes in both direct and indirect intermolecular interactions in the excited-state manifold. With decreasing thickness, the excitonic absorption spectrum blueshifts, the DS energy diminishes, the Stokes shift increases, and signatures of enhanced exciton delocalization emerge. These results, largely consistent with the Frenkel-CT mixing framework, show that dimensional reduction offers distinct ways to control molecular excitons, establishing the layer number as a new variable for tuning intermolecular excited-state interactions in molecular crystals.

## Results and Discussion

***Thickness-defined 2D molecular crystals by mechanical exfoliation.*** Molecular crystals can be mechanically fragile and structurally prone to disorder during thinning. We examined whether high-quality molecular crystals could be mechanically exfoliated into thickness-defined 2D forms while preserving their pristine packing structure. Mechanical exfoliation[16] should be most effective in crystals with strongly anisotropic intermolecular cohesion (Fig. 1a), where robust in-plane interactions (within the ***x-y*** plane) coexist with comparatively weak out-of-plane binding along the ***z*** axis. Tc crystals satisfy this requirement: Their herringbone lattice with two basis molecules on the (001) facet, or ***a-b*** plane in Fig. 1b, forms an extended π-stacked network, constituting a single-layer (1L) Tc. In contrast, adjacent Tc layers are connected primarily through weaker peripheral C–H contacts (Fig. 1c). Density functional theory calculations on a 1L-slab in Fig. 1c support this structural picture: Among the 4 low-index facets considered (Fig. S1 and Methods), cleavage along the (001) plane exhibits the lowest cleavage energy[31], substantially lower than competing crystallographic directions (Fig. 1d). The (001) energy is ~50% smaller than the cleavage energy of graphite[32], placing mechanical isolation of 2D Tc within the range of established exfoliable materials[33]. We also note that the interlayer binding energy calculated without lattice relaxation was 5–15% smaller in a 2L-slab model (Fig. S1), indicating a modest sensitivity of the calculated separation energy to structural relaxation and slab configuration.

Guided by this analysis, we developed a mechanical exfoliation procedure (Methods) for molecular crystals (see Fig. S2 for recrystallization). Atomic force microscopy (AFM) revealed laterally extended terraces with nanometer-scale flatness over micron-scale distances (Fig. 1e). The minimum step height matched the interplanar spacing of the bulk crystal (Fig. 1f, top), consistent with exfoliation in discrete molecular layers (Fig. 1f, bottom). Few-layer regions down to 2L were routinely obtained across multiple substrates (Si wafers, quartz and sapphire). For rapid thickness identification in Tc, we established substrate-specific optical-contrast calibrations spanning tens of layers (Fig. 1g): The optical contrast, defined in Methods, scaled monotonically with thickness and enabled non-destructive layer counting prior to spectroscopy. To test the broader scope of this strategy, we applied the same procedure to three additional molecular semiconductors: Pc, DIP, and PTZ. For each crystal with anisotropic bonding motifs, ultrathin flakes were successfully isolated (Fig. S3; see also Fig. 1e for Pc and DIP), indicating that exfoliation is not unique to Tc but is accessible across a wider class of quasi-layered molecular solids. Notably, 1L Pc in Fig. 1e could be identified with topographic analysis (Fig. S3). These results show that mechanical exfoliation yields molecular crystals with layer number defined at near-molecular precision and reaches the mono- or few-layer regime, without chemical processing or epitaxial templating.

***Structural integrity and crystal packing of 2D Tc.*** Any interpretation of thickness-dependent excitonic behavior requires establishing that exfoliation does not introduce significant disorder, mosaicity (domain misorientation) or unintended structural phase changes. We therefore examined the crystalline integrity of few-layer Tc using polarization-resolved photoluminescence (PL) imaging and electron diffraction. The former, implemented in a wide-field format (Methods), exploits the fact that the lowest Tc emission is polarized along a crystallographically defined in-plane axis, allowing molecular orientation to be mapped optically across extended flakes[8]. As depicted in Fig. 2a, the lowest electronic transition of an isolated Tc molecule splits into a doublet of lower and upper Davydov states (LDS and UDS) in bulk[5] and 2D[8] crystals with two basis molecules. Polarized PL spectra of exfoliated 11L Tc (Fig. 2b) mainly consist of two vibronic transitions from LDS, 0-0 and 0-1, exhibiting dipolar character aligned along the ***b*** axis[8] (inset in Fig. 2b).

Wide-field polarimetric imaging of a multi-terraced sample (Fig. 2c) revealed uniform emission anisotropy across several thickness domains: Two images of polarization-resolved PL intensities, horizontal ($I_H$ in Fig. 2d) and vertical ($I_V$ in Fig. 2e), lead to the orientation ($\theta_b$) image of the sample (Fig. 2f). $\theta_b$, representing the azimuthal angle of the ***b*** axis with respect to the laboratory ***y*** axis (inset of Fig. 2f), can be approximated as the following equation[8]: $\theta_b \cong \pm\tan^{-1}[(I_H/I_V)^{1/2}]$, the accuracy and noise-dependence of which were systematically verified (Figs. S4 and S5). Despite stepwise variations in thickness (5–20L), the in-plane orientation remained constant across the entire flake (Fig. 2f), indicating that exfoliation preserves single-crystal registry during cleavage, not producing misoriented fragments. The narrow angular distribution (FWHM = 1.9°), comparable to the ~1.9° effective angular resolution at high signal-to-noise ratios, places an upper bound on the orientational disorder within the measured region (Figs. S4 and S5).

Selected-area electron diffraction (SAED) independently confirmed long-range order. The five-layer exfoliated Tc crystal in Fig. 2g displayed sharp reciprocal-lattice features indexed to the known high-temperature polymorph of bulk Tc[34]. As summarized in Table 1, the mean unit-cell parameters of exfoliated Tc crystals closely matched the bulk reference values over the measured thickness range of 5–27L. Therefore, mechanical thinning to few-layer flakes does not induce substantial lattice reconstruction. This structural fidelity is a key distinction from many deposited organic thin films[35], where strain[36], substrate templating[8, 21] or kinetically trapped polymorphs[37] can obscure intrinsic dimensional trends. By contrast, assembly-grown ultrathin Tc on graphene exhibited a measurable in-plane lattice contraction (~4% along the ***a*** axis) unlike exfoliated crystals (Table 1). Although modest in magnitude, such packing changes are expected to strongly influence intermolecular electronic couplings that govern excitonic states because Coulomb and exchange interactions depend sensitively on intermolecular separation and slip geometry[4]. This comparison helps distinguish the effects of molecular packing from those of thickness on the excitonic states as discussed below.

***Excitonic manifold reconstructed by packing and dimensionality.*** We next examined how out-of-plane confinement modifies the elementary excitonic structure of 2D Tc. We adopt the Frenkel-CT mixing framework by Spano and coworkers[4], in which the two Davydov-split excited states of bulk Tc crystals arise from the interplay between direct Coulomb and CT-mediated couplings (Fig. 2a). In the underlying molecular configurations, the electron and hole occupy the same molecule for a Frenkel excitation and neighboring molecules for a CT excitation. The direct Coulomb coupling contributing to Frenkel excitons arises from electrostatic interactions between molecular transition densities. By contrast, the Frenkel-CT coupling matrix elements are governed primarily by intermolecular electron and hole transfer integrals, which depend on molecular orbital overlap. We use this framework to examine how thickness-dependent dielectric screening[13] and molecular packing change the diabatic energies and intermolecular coupling (Supplementary Note). As shown in Fig. 3a, polarization-resolved absorption spectroscopy via differential reflectance (DR)[38] resolved the lower and upper Davydov branches together with their vibronic progressions in exfoliated 2D Tc. The orthogonal polarization selection rules[8] of the two branches were retained (Fig. 3b), indicating that the basic Davydov-split excitonic structure survives in the 2D limit. As shown in Fig. 3c, however, the energy separation between the two branches, DS energy, decreased systematically with decreasing thickness (see Fig. S6 for the changes in the energies of LDS and UDS). Assembly-grown crystals showed an even steeper reduction in DS energy as multilayers thinned to the monolayer regime and exhibited enhanced splitting compared to exfoliated systems. Thus, the DS energy varies sensitively with both molecular packing and crystal thickness.

In Fig. 2a, we denote the predominantly ***b***- and ***a***-polarized Frenkel states before CT mixing by $\mathrm{F_b^o}$ and $\mathrm{F_a^o}$, with energies $\mathrm{E_{F,b}^o}$ and $\mathrm{E_{F,a}^o}$, respectively. These energies include local excitation energies and contributions of Coulomb couplings across the crystal lattice. The $\mathrm{F_b^o}$ state lies slightly higher than the $\mathrm{F_a^o}$ state, before CT mixing, but mixes with a higher-lying CT state with energy $\mathrm{E_{CT}}$ through a CT coupling V to form the LDS. On the other hand, as explained by Yamagata et al.[5], $\mathrm{F_a^o}$ remains as the unmixed UDS within the idealized monoclinic limit that assumes equal electron and hole dissociation integrals ($\mathrm{D_e = D_h}$) and neglects vibronic coupling. For actual triclinic Tc, these labels specify the predominant polarization directions of the optical transitions and are irrelevant to symmetric or antisymmetric Frenkel eigenstates.

First, the DS energy difference between exfoliated and assembly-grown crystals can be associated with lattice contraction in the latter, as observed in diffraction patterns. Denser packing can strengthen Coulomb interactions. Because the leading-order term of the pairwise Coulomb coupling is dipolar, its magnitude scales inversely with the cube of the intermolecular distance (R). The estimated 3.2% and 1.2% decreases in $R_+$ and $R_-$ (Fig. S7), calculated from the mean in-plane unit-cell parameters (Table 1), suggest ~10% and ~4% increases in the magnitudes of the corresponding pairwise couplings, respectively, assuming comparable dielectric environments and unchanged transition-dipole magnitudes and orientation factors. Using the unscreened reference couplings reported by Yamagata et al.[5], we estimate a 6–7 meV increase in the contribution of these nearest-neighbor pairs to the diabatic Frenkel separation $\Delta_F^o = E_{F,b}^o - E_{F,a}^o$, before dielectric screening (Supplementary Note, Eqs. S3–S5). These positive contributions tend to widen the diabatic separation and lower the ***a***-polarized Frenkel state relative to the mean Frenkel energy.

The polarized DR spectra in Fig. 3d and 3e show that all the vibronic transitions associated with both the LDS and UDS directions are redshifted in the grown crystals. Notably, the 0-0 energy difference of LDS between the two crystal types was greater than that of UDS, resulting in larger DS energies for the grown crystals (Fig. 3c). For 5L crystals, for example, grown crystals exhibited a splitting ~15 meV larger than that of exfoliated ones (~79 meV). Denser packing can additionally lower the diabatic CT-state energies through stronger electron-hole attraction and polarization stabilization, and enhance Frenkel-CT coupling through increased intermolecular orbital overlap. These changes can strengthen Frenkel-CT mixing and increase the stabilization of the LDS, denoted by $\delta_{CT}$ (Fig. 2a). Stronger net negative same-sublattice Coulomb coupling can also lower the mean Frenkel energy, shifting both diabatic Frenkel states downward. Relative to this mean, a larger $\Delta_F^o$ lowers $E_{F,a}^o$ and raises $E_{F,b}^o$. Enhanced CT-induced stabilization can then give the LDS a larger redshift than the UDS, provided that the increase in $\delta_{CT}$ exceeds that in $\Delta_F^o$. This scenario accounts for the redshifts of both transitions and the larger DS energy observed in the grown crystals (Fig. S8; Supplementary Note, Eq. S2b).

Second, despite their different packing, both crystal classes showed decreasing DS energy upon thinning. In exfoliated Tc, both the LDS and UDS 0-0 absorption energies shift to higher energies with decreasing thickness, with a larger blueshift of the LDS (Fig. S6). These observations strongly suggest that reduced dimensionality is a primary contributor to the thickness dependence, beyond packing variation alone. A natural microscopic origin is the evolution of dielectric screening in the few-layer limit[13]. The local dielectric response relevant to pairwise Coulomb coupling is assumed to be less sensitive to dimensional reduction than the response governing CT-state stabilization[39, 40].

On the basis of the analysis above, we can focus on dielectric-induced energy shifts, employing a common-shift approximation in which the two diabatic Frenkel states shift by the same amount, leaving $\Delta_{\mathrm{F}}^{\mathrm{o}}$ unchanged (see the Supplementary Note for the underlying assumptions). The Frenkel-CT coupling V is also held fixed as a simplifying assumption based on the nearly bulk-like in-plane lattice parameters (Table 1). In this limit, the observed UDS blueshift tracks an upward shift of $\mathrm{E}_{\mathrm{F,a}}^{\mathrm{o}}$, whereas the mixed LDS responds to both $\mathrm{E}_{\mathrm{F,b}}^{\mathrm{o}}$ and $\mathrm{E}_{\mathrm{CT}}$. The observed UDS blueshift sets the common Frenkel upshift phenomenologically. If all three diabatic energies shifted upward by the same amount, the DS energy would remain unchanged. Within the idealized model, the larger LDS blueshift is reproduced when the CT state shifts upward more than the two diabatic Frenkel states (Fig. S9a; Supplementary Note, Eqs. S8a and S8b). Decreasing thickness, which weakens dielectric screening, can reduce the polarization stabilization of the separated charges in a CT configuration. Weaker screening also strengthens their mutual attraction, but the CT energy rises when the loss of ionic polarization stabilization exceeds this opposing effect ($\mathrm{r_C} > \mathrm{r_P}$ in the simplified Born model; Supplementary Note, Eq. S7; Fig. S9b). This relative CT-state destabilization would weaken Frenkel-CT hybridization and reduce the CT-induced stabilization of the LDS, increasing its Frenkel character and providing its additional blueshift. The observed Davydov-state energy trends and their qualitative interpretation within the Frenkel-CT mixing model are summarized in Table 2.

***Confinement-enhanced excited-state relaxation and delocalization.*** Beyond the absorption manifold, reduced dimensionality also strongly modifies excited-state relaxation. We compared the lowest vibronic transition energies in absorption and emission as a function of thickness and found that both shifted to higher energies upon thinning (Fig. 4a). However, the energy difference between them (Stokes shift) increased markedly toward the few-layer limit, reaching approximately three times its bulk-like value in the thinnest crystals.

Such behavior is consistent with more pronounced structural and environmental relaxation following photoexcitation in ultrathin crystals, although thickness-dependent defect trapping or substrate interactions may additionally influence the measured shift. In bulk molecular solids, lattice rigidity suppresses large-amplitude nuclear reorganization. By contrast, few-layer crystals possess free surfaces, reduced out-of-plane constraint and enhanced susceptibility to substrate-induced corrugation or strain[41]. These factors can increase relaxation along low-frequency intermolecular coordinates coupled to the excited states, allowing neighboring molecules to adjust their relative positions more readily and thereby increasing the reorganization energy. Stokes shifts of large excitonic polarons are predicted to increase as the size of one-dimensional molecular aggregates decreases due to enhanced electron-phonon coupling[42].

One hallmark of collective excited states is their spatial coherence. To probe exciton delocalization, we analyzed the PL intensity ratio between the 0-0 and 0-1 vibronic transitions (Fig. 4b). In molecular aggregates, enhanced 0-0 emission is a characteristic signature of coherent excitonic states because the zero-vibration transition undergoes superradiance unlike vibronic sidebands[4]. The 0-0/0-1 ratio increased steadily as thickness decreased, approximately doubling in intermediate multilayers and rising further in the few-layer regime. Within Spano's conventional framework, in which the PL ratio serves as a spectroscopic measure of exciton coherence[4], the observed trend is consistent with progressively increasing exciton delocalization as the crystal is thinned, whereas thickness-dependent changes in vibronic coupling may also affect the absolute ratio. Notably, exfoliated and assembly-grown samples exhibited similar thickness dependence despite their different absolute transition energies, suggesting that coherence enhancement is governed primarily by dimensional confinement rather than by packing differences.

Greater delocalization in thinner crystals may be facilitated by changes in the balance between direct Coulomb coupling and CT-mediated couplings under reduced screening. As the dielectric screening decreases, dipole-dipole interactions can be enhanced, which may favor the extension of Frenkel excitations over multiple molecules. The net effective coupling also depends on the signs and magnitudes of the CT-mediated contributions and their interference with the direct Coulomb coupling. In addition, within the common-shift approximation above, our analysis suggests that reduced screening increases the energetic separation between the ***b***-polarized diabatic

Frenkel state and the relevant CT state, thereby reducing their hybridization and increasing the Frenkel character of the LDS, consistent with the simultaneous blueshifts and reduced DS energy (Figs. 3c, S6 and S9a; Supplementary Note). The increased Frenkel character may further favor spatial coherence when effective Frenkel coupling is sufficiently strong relative to disorder and exciton-phonon interactions. Under these conditions, local exciton-phonon interactions can be effectively averaged over multiple molecules. In this regard, the enhanced coherence inferred from the PL ratio could also be accompanied by greater superradiant character, as reported for delocalized excitons in tetracene[7]. Thus, the enhanced Stokes shift and increased exciton delocalization can coexist without necessarily sharing a common microscopic origin: The former is consistent with enhanced intermolecular structural relaxation, whereas the latter may reflect distinct changes in effective intermolecular coupling under reduced dielectric screening.

Thickness-controlled coherence has important functional implications because exciton size influences transport[43], radiative rates[7] and multi-exciton processes such as singlet fission[30]. The thickness-dependent changes in DS and Stokes shift and the PL intensity ratio suggest that dimensional reduction modifies intermolecular coupling, excited-state relaxation and exciton coherence through distinct microscopic processes in molecular crystals. Exfoliated 2D molecular solids therefore provide a complementary excitonic platform to inorganic 2D semiconductors, but in the regime of tightly bound and highly collective molecular excitations.

## Conclusion

We demonstrated that molecular crystals can be mechanically exfoliated into structurally well-defined 2D forms suitable for probing collective excited states. Electron diffraction and polarization-resolved optical measurements confirm that crystalline order and optical anisotropy are retained upon exfoliation. Using tetracene as a model system, we showed that reducing thickness toward the few-layer limit continuously modulates the excitonic energy landscape while largely preserving the main features of the bulk crystal structure. The structural fidelity of exfoliated crystals allows thickness-dependent effects to be examined with minimal changes in molecular packing, in contrast to the structural variations often encountered in deposited organic thin films. Out-of-plane confinement reduces the DS energy, substantially enhances the Stokes shift and increases the spectroscopic signatures of exciton delocalization, demonstrating that intermolecular coupling, excited-state relaxation and coherence evolve systematically with layer number. Extending this approach to other molecular crystals can enable systematic studies of dimensional effects from the bulk to the few-layer and monolayer limits, including their influence on singlet fission, exciton transport, polaritonic interactions and correlated excited states. Thus, this work establishes the layer number of molecular crystals as a key variable for controlling the nature of collective excited states, complementing molecular structure, packing and interfacial design.

## Methods

***Top-down preparation of samples.*** Before mechanical exfoliation, quartz (qz) and sapphire (sph) slides were cleaned with piranha solution and treated with a UV-ozone (UVO) cleaner (Jaesung Eng., UVC-30). The dry-cleaned substrates were subsequently annealed at 350 °C to remove residual surface charges. Silicon wafers with 285-nm-thick $SiO_2$ were UVO-cleaned and annealed at 300 °C.

Bulk powders of tetracene (TCI Co., > 97.0%), pentacene (TCI Co.), phenothiazine (TCI Co., > 99.9%) and diindenoperylene (Ambeed Inc., > 97.0%) were recrystallized using the physical vapor transport (PVT) method[44] to improve purity and obtain larger crystalline domains. During PVT, Ar gas was introduced into a quartz tube containing source powders (upstream) and Si substrates in an inner tube (downstream) at a flow rate of 50–200 mL/min. The temperatures of the source and recrystallization zones were independently controlled using a three-zone furnace. The former temperature was set to ensure sublimation of source powders, whereas the latter was optimized to achieve maximal deposition within the recrystallization zone (Fig. S2). Growth time was varied from a few to ~24 h. Crystalline flakes of molecular solids (typically several mm across) were grown on either the inner wall of the tube or the substrates. Typically, recrystallized crystals had flat areas that were 2–3 orders of magnitude larger than those of the as-purchased powders. Flakes with domain sizes larger than ~500 μm were used for mechanical exfoliation.

Because of their fragility relative to inorganic layered materials such as graphite and hexagonal BN (hBN), the molecular crystals were exfoliated using Scotch tape with minimal repeated thinning to avoid fracture. The tape-supported crystals were stamped onto the target substrate, gently rubbed to promote contact and left in contact under ambient conditions for several hours to several days to improve the exfoliation yield. The tape was then peeled off very slowly to minimize mechanical damage.

***Bottom-up preparation of samples.*** Few-layered Tc crystals were also prepared by assembly-assisted growth using a conventional thermal evaporator[8]. The deposition chamber was maintained below $2 \times 10^{-6}$ Torr by a turbomolecular pump during deposition. Exfoliated few-layer graphene and few-nm hBN were used as assembly templates. For growth, the templates were exposed to Tc vapor at a nominal deposition rate of 0.1–0.5 Å/s. The integrated flux corresponded to a nominal film thickness of 5–10 nm. The templates were maintained at 40 °C to induce ordered assembly growth.

***AFM measurements.*** Sample thickness was determined by atomic force microscopy (AFM). Height images were acquired with an AFM system (Park Systems, XE-70) operated in non-contact mode under ambient conditions.

***Electron diffraction measurements.*** Selected-area electron diffraction (SAED) patterns were obtained using a transmission electron microscope (JEOL, JEM-2100F) operated at an accelerating voltage of 80 kV (selected-area aperture diameter, ~1.2 μm) at the Materials Imaging & Analysis Center of POSTECH. Diffraction signals were recorded using a CCD camera (Gatan, Orius SC200) with exposure times of 1–4 s to minimize electron-beam-induced damage. To further protect the samples from electron-beam damage, both sample types were encapsulated between single- or few-layer graphene flakes. Bottom-up samples grown on graphene were capped with another graphene layer using the dry transfer method[9]. Top-down samples were directly exfoliated onto pre-exfoliated graphene on $SiO_2$/Si substrates and immediately capped with a second graphene layer by dry transfer. To improve the yield of this double-exfoliation process, $SiO_2$/Si substrates were treated with low-power air plasma for 5 min, followed by mechanical exfoliation of the first graphene layer while the substrates were maintained on a hot plate at 100 °C[45]. The graphene-sandwiched Tc crystals were transferred onto silicon nitride TEM grids (Norcada, NH050D2) using an ethanol-assisted wet transfer method[9]. Briefly, a thin PDMS film (~1 mm thick) was attached to the sandwiched sample region, and the assembly was immersed in ethanol for 2–6 h to transfer the sample to the PDMS. The PDMS-supported sample was then dry-transferred to a silicon nitride grid pre-cleaned by $O_2$ or air plasma for 10 min.

***Differential reflectance and optical contrast measurements.*** Differential reflectance (DR) spectroscopy was used to probe the optical absorption of 2D molecular crystals supported on transparent substrates. The DR spectra were acquired using a collimated tungsten-halogen lamp as a broadband source. The incident beam was focused onto the sample, and reflected signals were collected in a backscattering geometry using a 40× microscope objective (NA = 0.6) and directed to a spectrometer equipped with a thermoelectrically cooled CCD detector (Andor Inc., DU971P).

The typical probe diameter was ~2 μm. We acquired all DR spectra in the parallel-polarization configuration, aligning the polarizations of the incident and reflected light using linear polarizers. Samples were rotated using a rotational mount with a precision of 0.2°. For optically thin films supported on transparent substrates, in the weak-optical-response limit, DR (defined as $\frac{R_{sam}-R_{sub}}{R_{sub}}$) can be approximated by $DR \approx \frac{4}{n_{sub}^2-1}A$[38], where $R_{sam}$ and $R_{sub}$ are the reflected intensities of the sample-covered and bare-substrate regions, respectively. $n_{sub}$ and $A$ are the refractive index of the substrate and the absorptance of the sample, respectively. The optical contrast (OC) of the samples was used to determine their thicknesses and was defined analogously, using green-channel data extracted from optical micrographs.

***Photoluminescence spectroscopy.*** The homebuilt photoluminescence (PL) spectroscopy setup has been described previously[8]. Briefly, a solid-state laser operated at 457.8 nm (Cobolt, Twist) was focused onto the samples to a spot size of ~1 μm using a microscope objective (40×, NA = 0.6). Backscattered PL signals were collected using a CCD camera (Princeton Instruments, PyLon) connected to a spectrometer (Princeton Instruments, SP2300). All measurements were performed under ambient conditions. The PL intensity ratio was determined from the integrated areas of the fitted 0-0 and 0-1 peaks.

***Wide-field PL imaging.*** PL imaging was performed using the PL spectroscopy setup described previously[8, 46]. Briefly, the linearly polarized excitation beam (457.8 nm) was focused onto the back focal plane of the objective using a plano-convex lens (focal length = 400 mm). The resulting wide-field excitation produced an illumination area with a FWHM of ~250 μm. PL images were projected onto the CCD camera after passing through a 500–550 nm bandpass filter. The imaging area of the CCD (100 × 100 pixels) corresponded to a sample area of 32 μm × 32 μm with the 40× objective. Two PL images were sequentially acquired with horizontal and vertical excitation polarizations to extract crystallographic orientation using ratiometric polarimetry[8]. An analyzing polarizer was placed before the detector to select PL polarized parallel to the excitation polarization. The polarization-dependent sensitivity of the imaging setup was calibrated using amorphous copper oxide films as a reference for randomly polarized emission.

***Density functional theory calculations.*** Cleavage and interlayer binding energies were calculated with the Vienna Ab Initio Simulation Package (VASP)[47] using the Perdew-Burke-Ernzerhof (PBE) functional[48], the projector augmented-wave (PAW) method[49], and the D3(BJ) dispersion correction[50, 51]. The experimental crystal structure of tetracene (CSD refcode TETCEN01)[52] was used as the initial structure and relaxed until the residual forces were below 0.01 eV $Å^{-1}$. A plane-wave cutoff energy of 750 eV was used for both bulk and slab calculations. Γ-centered k-point meshes of 5 × 4 × 2 were used for the bulk, and 5 × 3 × 1, 5 × 2 × 1, 4 × 2 × 1 and 2 × 3 × 1 were used for the (100), (010), (001) and (110) slabs, respectively.

For each crystallographic facet, a 1L-slab model (Fig. 1c) was constructed with a vacuum region of 22 Å along the surface-normal direction and structurally relaxed. Cleavage energies were evaluated from the energy difference between the equilibrium and separated slab configurations, normalized by the in-plane area[31]. The vacuum spacing was verified by convergence tests. Interlayer binding energies were additionally estimated using a 2L-slab model (Fig. S1e). The binding energy was obtained from the energy difference per in-plane area between a bilayer of bulk structure and the corresponding bilayer with increased interlayer separation, without further structural relaxation.

## ASSOCIATED CONTENT

### Supporting Information

Cleavage energy and interlayer binding energy calculations; recrystallization of molecular crystals by physical vapor transport (PVT); 2D molecular crystals prepared by mechanical exfoliation; validation of crystal-orientation mapping by wide-field PL imaging; Tc orientation distribution as a function of the signal-to-noise ratio (S/N) in wide-field PL imaging; diagonal intermolecular separations used to estimate packing-dependent Coulomb couplings in Tc; thickness dependence of the two 0-0 Davydov states in Tc; schematic interpretation of packing-dependent Davydov-state energies; schematic interpretation of thickness-dependent Davydov-state energies; Frenkel-charge-transfer (CT) mixing model.

## AUTHOR INFORMATION

**Corresponding author**

*E-mail: sunryu@postech.ac.kr

NOTES

**Conflicts of Interest**

The authors declare no competing financial interest.

**Data Availability Statement**

The data that support the findings of this study are available from the corresponding author upon reasonable request.

## ACKNOWLEDGMENTS

This work was supported by the National Research Foundation of Korea (NRF-RS-2024-00336324, NRF-RS-2024-00411134, 2021R1A6A1A10042944) and the Glocal University 30 project. This research was also supported by the Korea Basic Science Institute (National Research Facilities and Equipment Center) grant funded by the Ministry of Education (RS-2026-25539607). S.J.J. acknowledges primary support from the US Department of Energy, Office of Science, Office of Basic Energy Sciences (DE-SC0026114) and support from Korea Institute for Advanced Study (KIAS) through its KIAS Scholar program.

## FIGURES & CAPTIONS

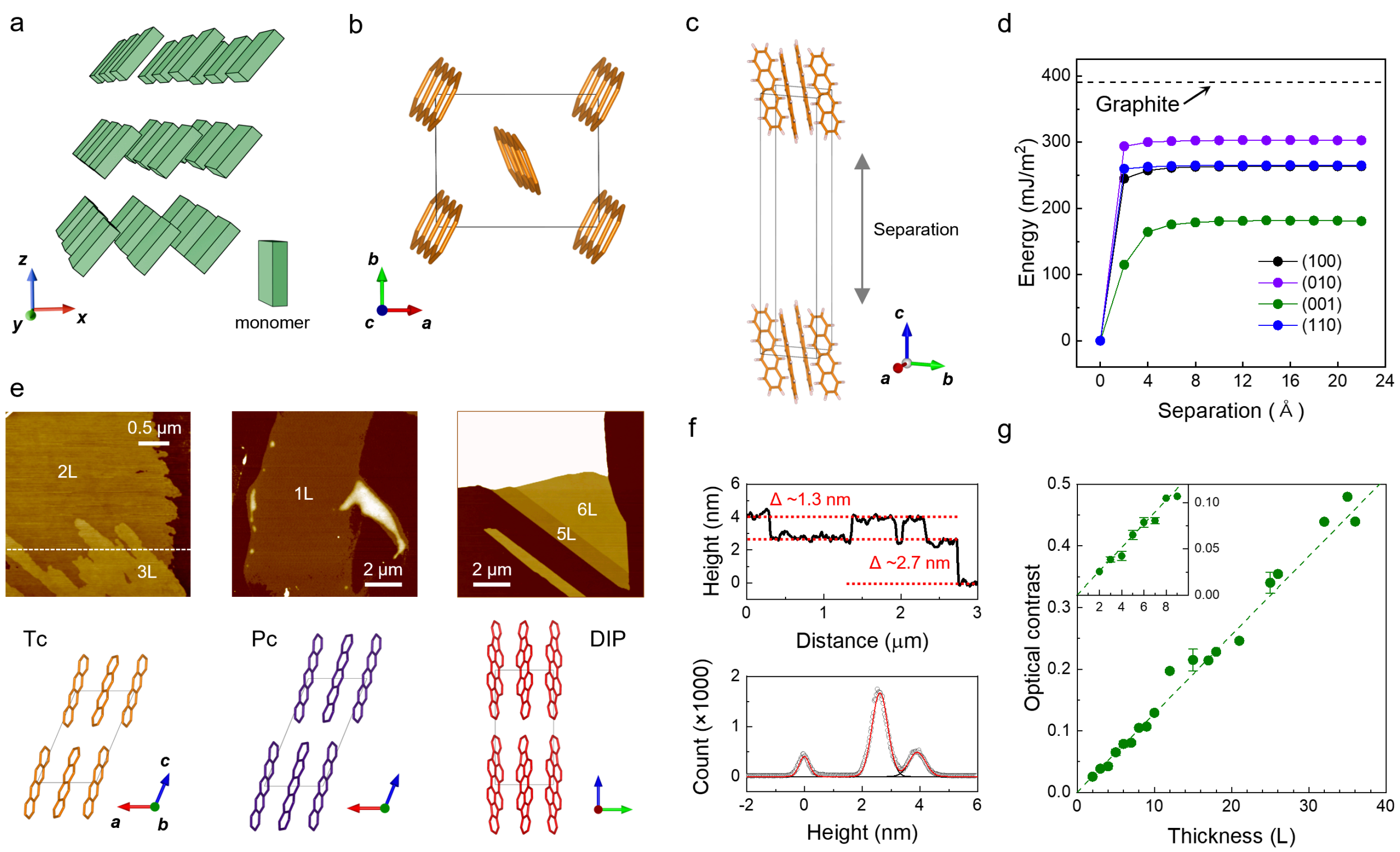

**Figure 1. Mechanical exfoliation of molecular crystals.** (a) Generic scheme of a molecular crystal structure. (b) Top view of 1L tetracene with unit cell (Tc). (c) Single-layer (1L) slab model of the (001) facet. (d) Energy of the 1L slabs for four facets as a function of interlayer separation, where the zero separation corresponds to the equilibrium state. The cleavage energy of each facet was defined as the energy difference between the equilibrium state and a separation of 22 Å. (e) AFM height images of exfoliated 2D Tc, pentacene (Pc) and diindenoperylene (DIP) crystals (top); side views of corresponding crystal structures (bottom), where gray solid lines represent their unit cells. (f) Height line profile across a one-layer step in Tc in (e), obtained along the dashed line (top); height histogram obtained from the entire AFM image (bottom). (g) Optical contrast as a function of thickness for exfoliated Tc samples supported on sapphire substrates. The inset shows an enlarged view of the data for thicknesses up to 6L.

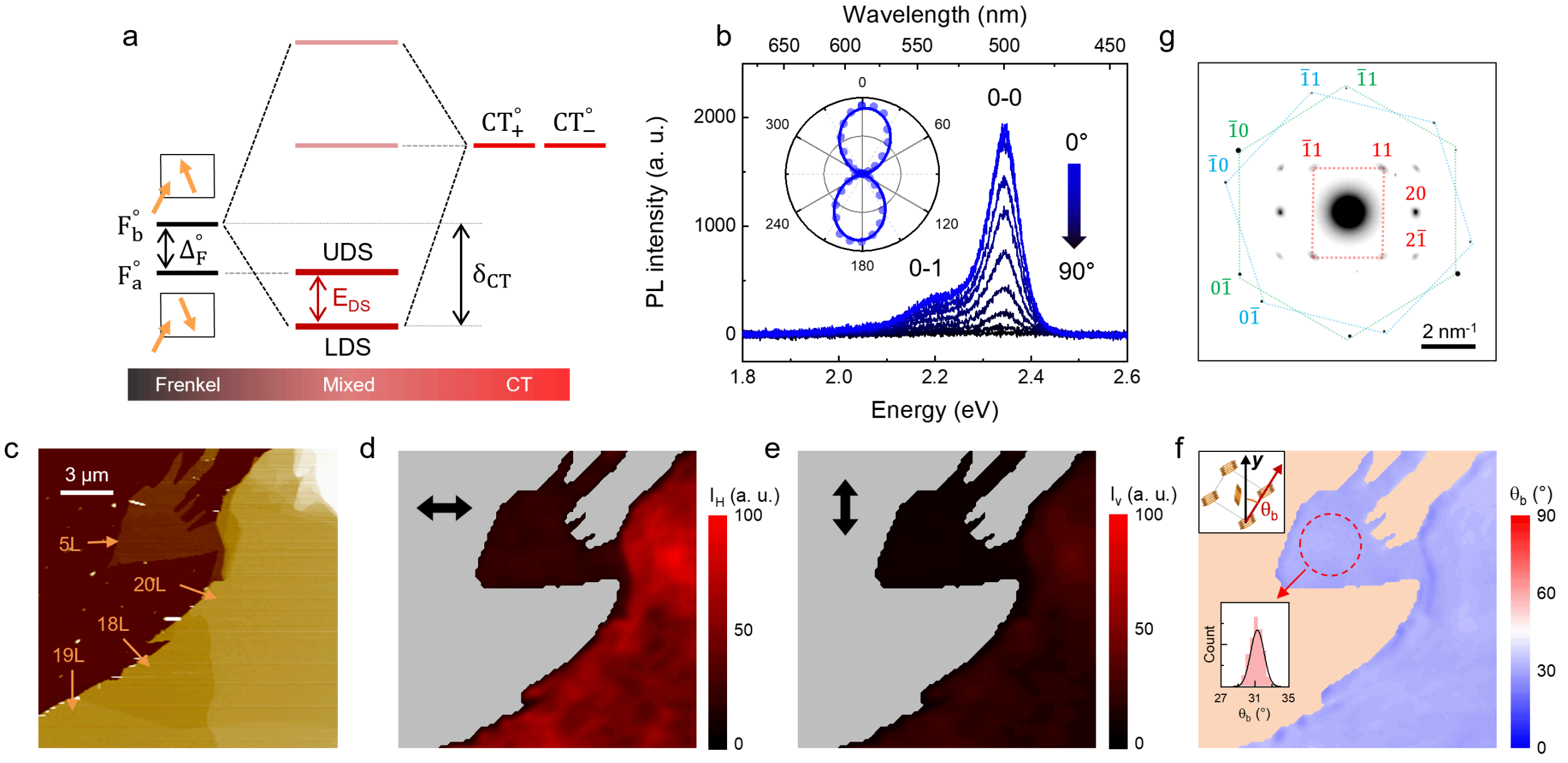

**Figure 2. Anisotropic optical properties and structure of exfoliated Tc.** (a) Minimal model of Frenkel-charge transfer (CT) mixing for herringbone-arranged Tc crystals (transition dipoles are denoted by orange arrows) in the idealized $D_e = D_h$ limit of Yamagata et al.[5]. Two diabatic Frenkel states ($F_b^o$ and $F_a^o$) represent the predominantly ***b***- and ***a***-polarized crystal excitations before CT mixing, with a separation $\Delta_F^o = E_{F,b}^o - E_{F,a}^o > 0$. $CT_+^o$ and $CT_-^o$ denote symmetric and antisymmetric CT combinations, respectively. In this model, $F_b^o$ mixes with $CT_+^o$ to form the lower Davydov state (LDS), whereas $F_a^o$ remains unmixed and forms the upper Davydov state (UDS). The CT-induced stabilization is $\delta_{CT} = E_{F,b}^o - E_{LDS}$, and the observed LDS-UDS splitting is $E_{DS} = E_{UDS} - E_{LDS} = \delta_{CT} - \Delta_F^o$ (Supplementary Note). The upper CT-mixed eigenstate lies $\delta_{CT}$ above $CT_\pm^o$. (b) Parallel-configuration polarized photoluminescence (PL) spectra of 11L Tc/sapphire obtained by rotating the sample in steps of 10°. Two prominent vibronic transitions are denoted as 0-0 and 0-1. The polar graph in the inset shows the 0-0 intensity as a function of rotation angle: The laser polarization is aligned approximately parallel to the ***b*** axis at 0°. (c) AFM height image of an exfoliated Tc sample with multiple terraces of various thicknesses. (d and e) Wide-field polarized PL images of the sample in (c), with horizontal ($I_H$ in d) and vertical ($I_V$ in e) polarization, as indicated by the black arrows. (f) Orientation ($\theta_b$) image obtained from (d) and (e), where $\theta_b$ is defined with respect to the laboratory ***y*** axis (inset). (g) Selected-area electron

diffraction (SAED) image obtained from graphene/5L Tc/graphene. Square and hexagonal dashed lines mark the diffraction of Tc and graphene crystals, respectively. Diffraction spots are indexed using 2D Miller indices (hk), corresponding to (hk0).

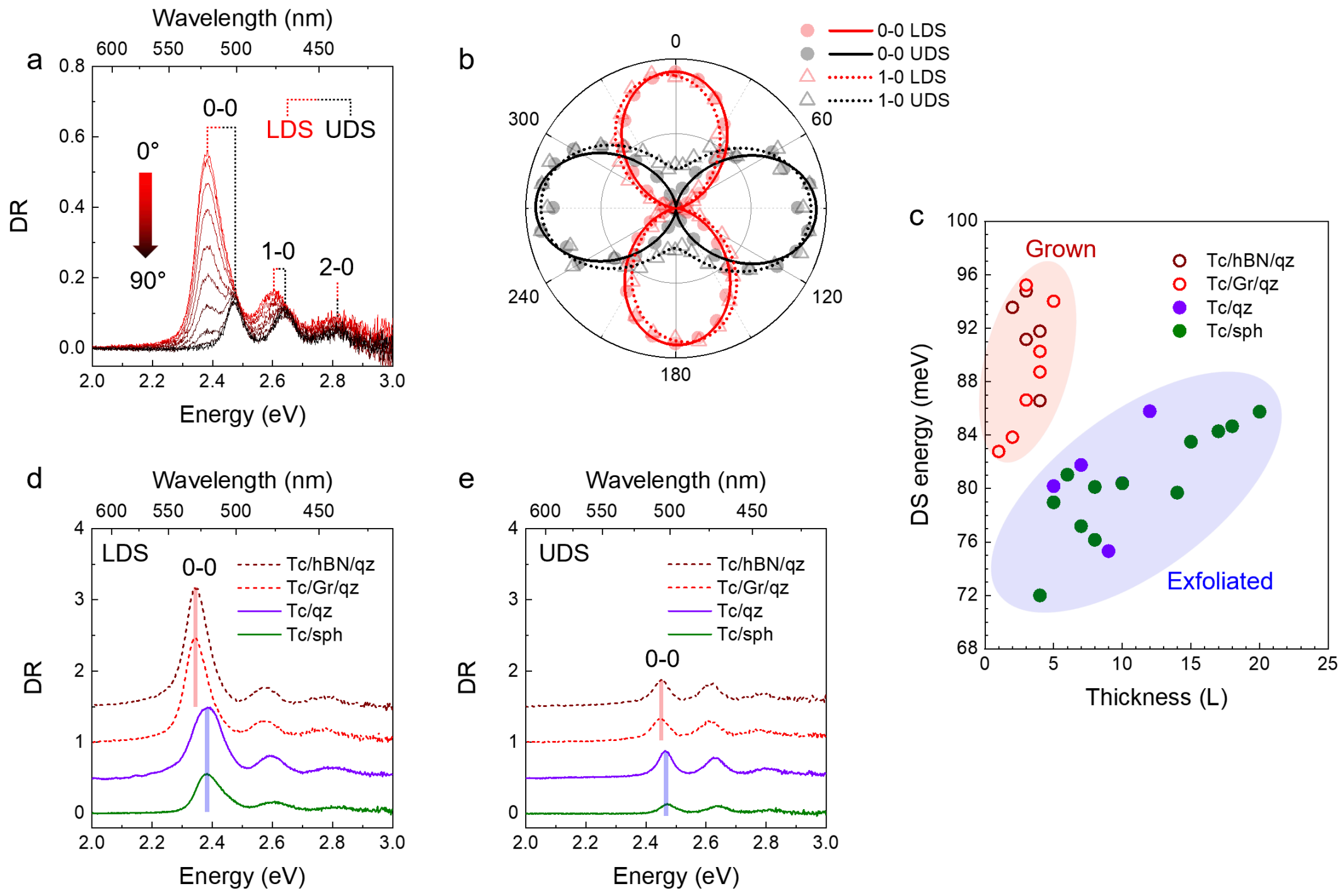

**Figure 3. Thickness-dependent Davydov splitting (DS) of exfoliated and grown Tc.** (a) Differential reflectance (DR) spectra of 5L Tc/sapphire obtained by rotating the sample in steps of 10°. Each vibronic transition is denoted as v-0, where v is the vibrational quantum number of the excited state. The Davydov splitting in the 2-0 peak could not be resolved. (b) Polar graph of normalized DR intensities for LDS and UDS transitions obtained from (a). (c) $E_{DS}$ of the 0-0 transition of exfoliated (filled) and grown (hollow) samples as a function of thickness. (d and e) Parallel-configuration DR spectra of exfoliated (Tc/qz and Tc/sph; solid lines) and grown (Tc/Gr/qz and Tc/hBN/qz; dashed lines) samples, polarized parallel to ***b*** for LDS (d) and ***a*** for UDS (e).

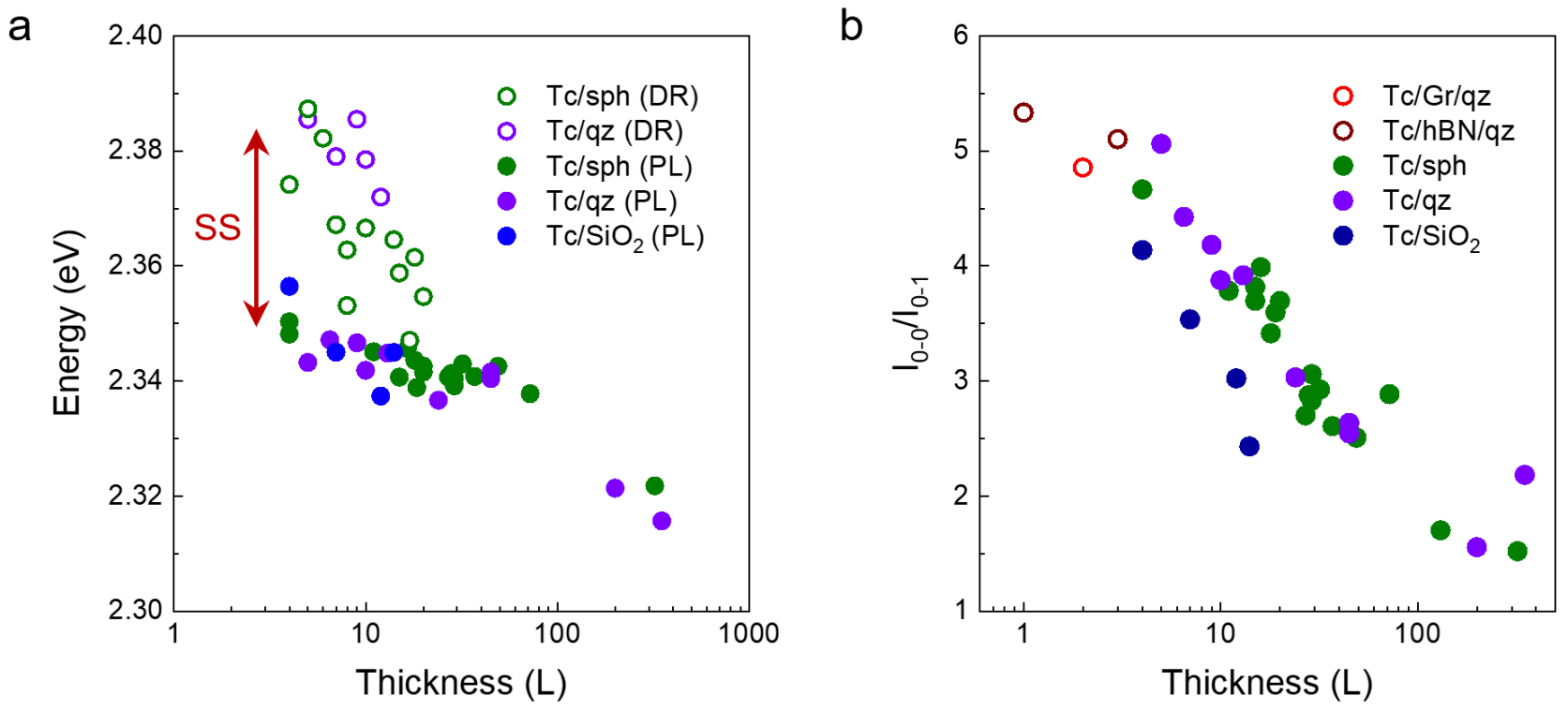

**Figure 4. Thickness-dependent Stokes shift and signatures of exciton delocalization.** (a) Absorption and emission energies (LDS of 0-0) of exfoliated samples as a function of thickness, obtained from DR (hollow) and PL (solid) measurements, respectively. The energy difference between the two measurements corresponds to the Stokes shift (SS, brown arrow). (b) PL intensity ratio of the 0-0 and 0-1 transitions ($I_{0-0}/I_{0-1}$) for exfoliated (filled) and grown (hollow) samples as a function of thickness. The transitions in (a) and (b) correspond to the LDS.

**TABLES**

| Samples | ***a*** (Å) | ***b*** (Å) | γ (°) | Notes |
|---|---|---|---|---|
| Assembly-grown Tc | 7.60(10) | 6.01(5) | 89.0(9) | 1–4L (15 samples) |
| | [-3.8%] | [-0.3%] | [+3.1%] | |
| Exfoliated Tc | 7.90(15) | 5.98(12) | 87.8(14) | 5–27L (6 samples) |
| | [0.0%] | [-0.8%] | [+1.7%] | |
| Bulk Tc | 7.90 | 6.03 | 86.3 | Ref. 34 |

**Table 1. Unit-cell parameters of 2D Tc crystals.** SAED patterns from multiple 2D samples were analyzed using graphene capping layers as internal standards. Standard deviations are given in parentheses. Relative deviations in square brackets are referenced to the bulk Tc values[34].

| Observable | Energy expression | Physical interpretation | Observed energy trends | |
|---|---|---|---|---|
| | | | Assembly-grown relative to exfoliated | Upon thinning |
| LDS energy ($E_{LDS}$) | $E_{LDS} = E^{o}_{F,b} - \delta_{CT}$ | Diabatic ***b***-polarized Frenkel energy lowered by Frenkel-CT mixing | Redshift exceeding that of the UDS | Blueshift exceeding that of the UDS |
| UDS energy ($E_{UDS}$) | $E_{UDS} = E^{o}_{F,a}$ | Unmixed ***a***-polarized diabatic Frenkel state | Redshift | Blueshift |
| Davydov splitting ($E_{DS}$) | $E_{DS} = E_{UDS} - E_{LDS} = \delta_{CT} - \Delta^{o}_{F}$ | Competition between CT stabilization and diabatic Frenkel separation | Increase | Decrease |

**Table 2. Observed Davydov-state energy trends and their qualitative interpretation within the Frenkel-CT mixing model.** The last two columns describe energy changes in assembly-grown crystals relative to exfoliated crystals at comparable thickness or upon thinning. Within each comparison, the LDS energy shift has a greater magnitude than the UDS shift. The individual-state blueshifts upon thinning refer to exfoliated Tc (Fig. S6); the decrease in Davydov splitting is observed in both crystal classes. The model relations adopt the unmixed-UDS approximation described in the Supplementary Note. The diabatic Frenkel energies include local excitation energies and lattice Coulomb contributions. Their common shift can also affect CT-induced stabilization through the Frenkel-CT energy gap.

**TOC Figure**

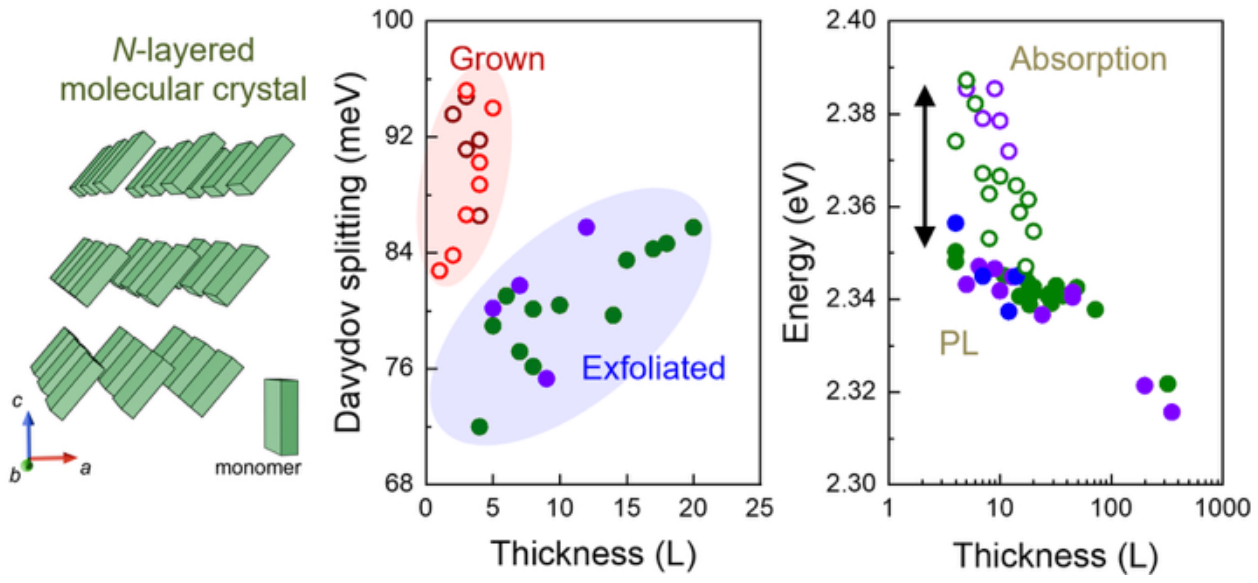

N-layered
molecular crystal
c
b
a
monomer
Grown
Exfoliated
Davydov splitting (meV)
Thickness (L)
Absorption
PL
Energy (eV)
Thickness (L)

Supplementary Information for

# Dimensional Control of Excitonic Interactions in Exfoliated 2D Molecular Crystals

Jonghyun Son, Seonghyun Koo, Daniel Yim, Sangjin Han, Dong-Hwan Yang, Gi-Yeop Kim, Kihyun Lee, Jieun Yeon, Hye Soo Kim, Eunbeen Jeon, Minji Ko, Minhee Choe, Kenji Watanabe, Takashi Taniguchi, Hee Cheul Choi, Kwanpyo Kim, Si-Young Choi, Seogjoo J. Jang, Hyungjun Kim and Sunmin Ryu*

*E-mail: sunryu@postech.ac.kr

**Contents**

## Supplementary Figures

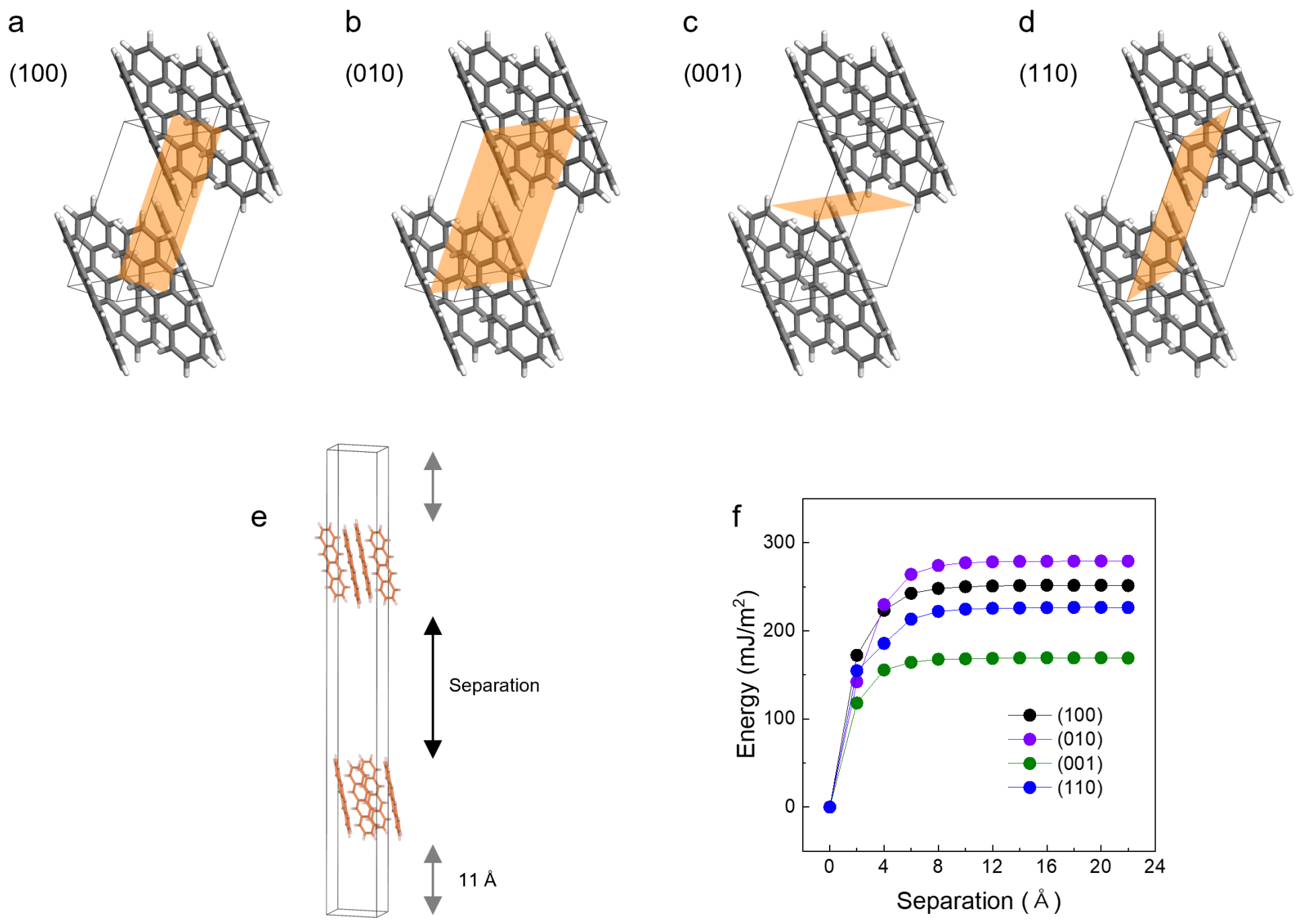

g

| Facets | Cleavage energy (mJ/m$^2$) (1L-slab with lattice relaxation) | Interlayer binding energy (mJ/m$^2$) (2L-slab without lattice relaxation) |
|---|---|---|
| (100) | 263 | 251 |
| (010) | 303 | 279 |
| (001) | 181 | 169 |
| (110) | 265 | 226 |

**Figure S1. Cleavage energy and interlayer binding energy calculations.** (a–d) Bulk Tc unit cell (black lines) and cleavage planes (orange) for the (100) (a), (010) (b), (001) (c) and (110) (d) facets. (e) 2L Tc slab model for the (001) facet with a total vacuum region of 22 Å (11 Å on each side). (f) Tc interlayer binding energy of the four facets as a function of separation. (g) Cleavage energy and interlayer binding energy for the four facets obtained from density functional theory calculations.

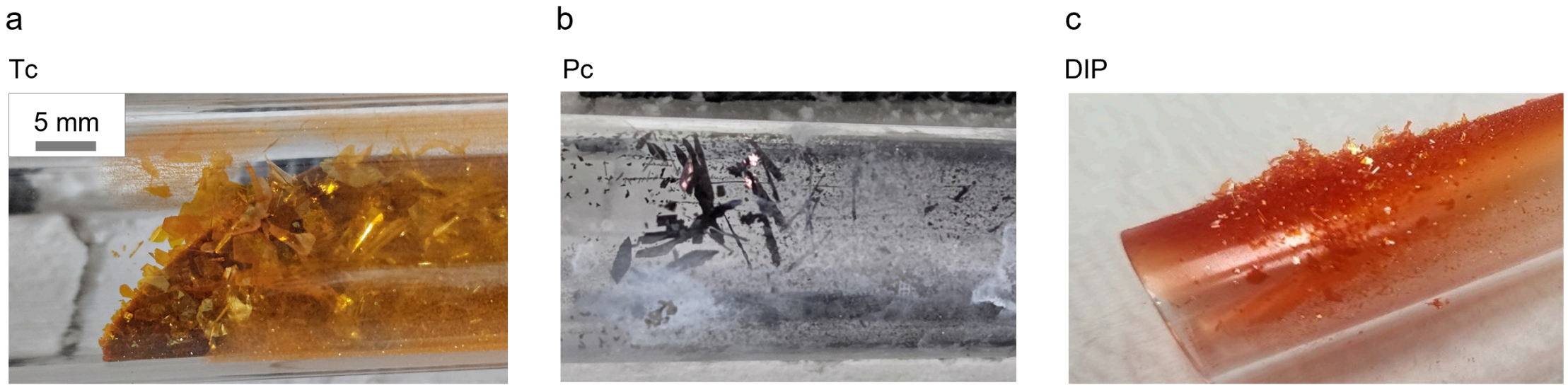

**Figure S2. Recrystallization of molecular crystals by physical vapor transport (PVT).** (a–c) Photographs of tetracene (Tc, a), pentacene (Pc, b) and diindenoperylene (DIP, c) recrystallized in quartz tubes.

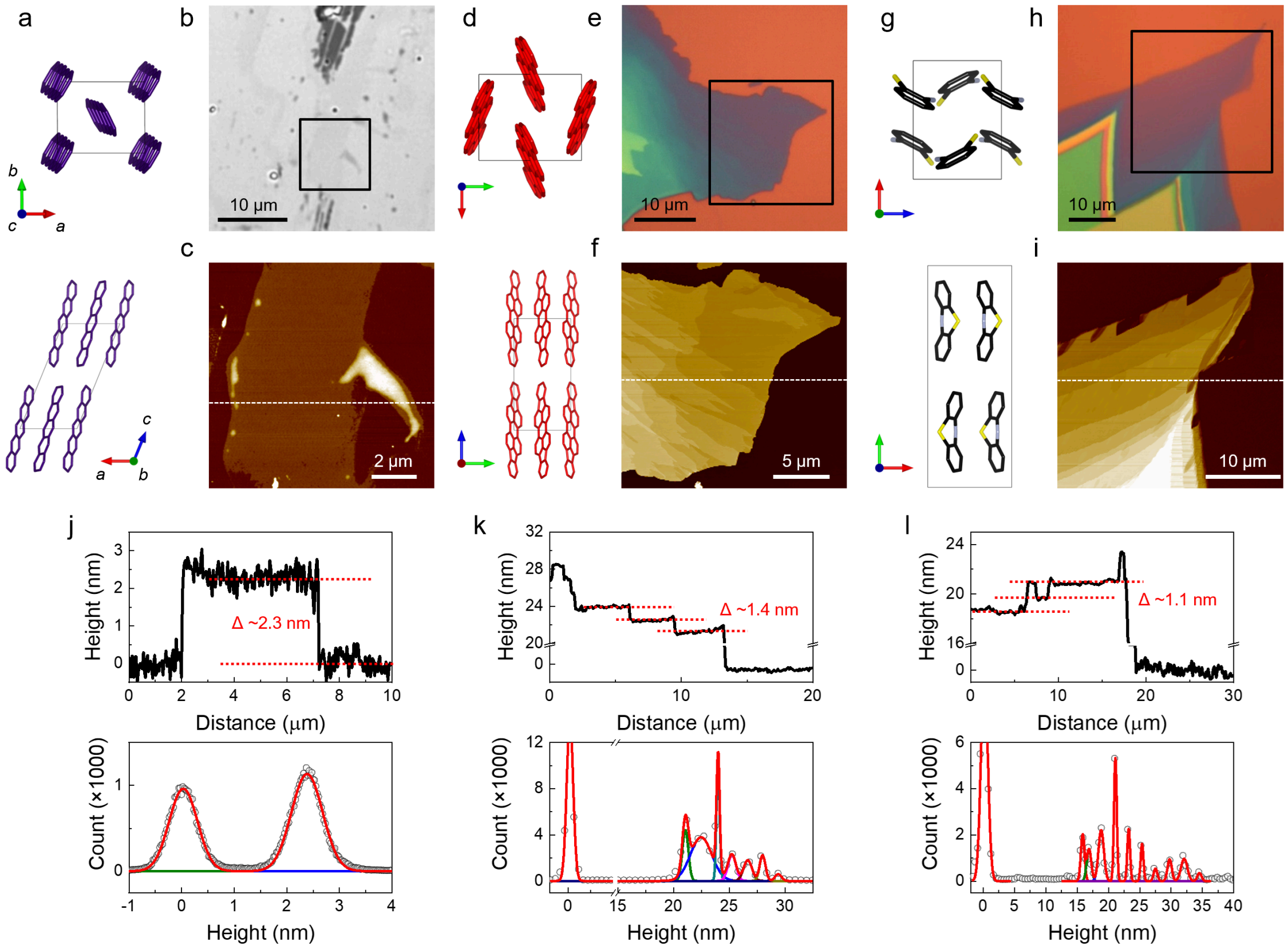

**Figure S3. 2D molecular crystals prepared by mechanical exfoliation.** (a,d,g) Top and side-view unit cells of the molecular crystal structures of Pc ($C_{22}H_{14}$) (a), DIP ($C_{32}H_{16}$) (d) and phenothiazine (PTZ, $C_{12}H_9NS$) (g). (b,e,h) Optical images of the exfoliated 2D crystals of Pc (b), DIP (e) and PTZ (h). (c,f,i) AFM height images corresponding to the black squares in (b), (e) and (h), respectively. (j,k,l) Height profiles along the dashed lines and the height histograms obtained from the AFM images in (c), (f) and (i), respectively. The step height of 1L Pc (2.3 nm) in (j) suggests an interfacial gap or an AFM height offset of ~0.8 nm.

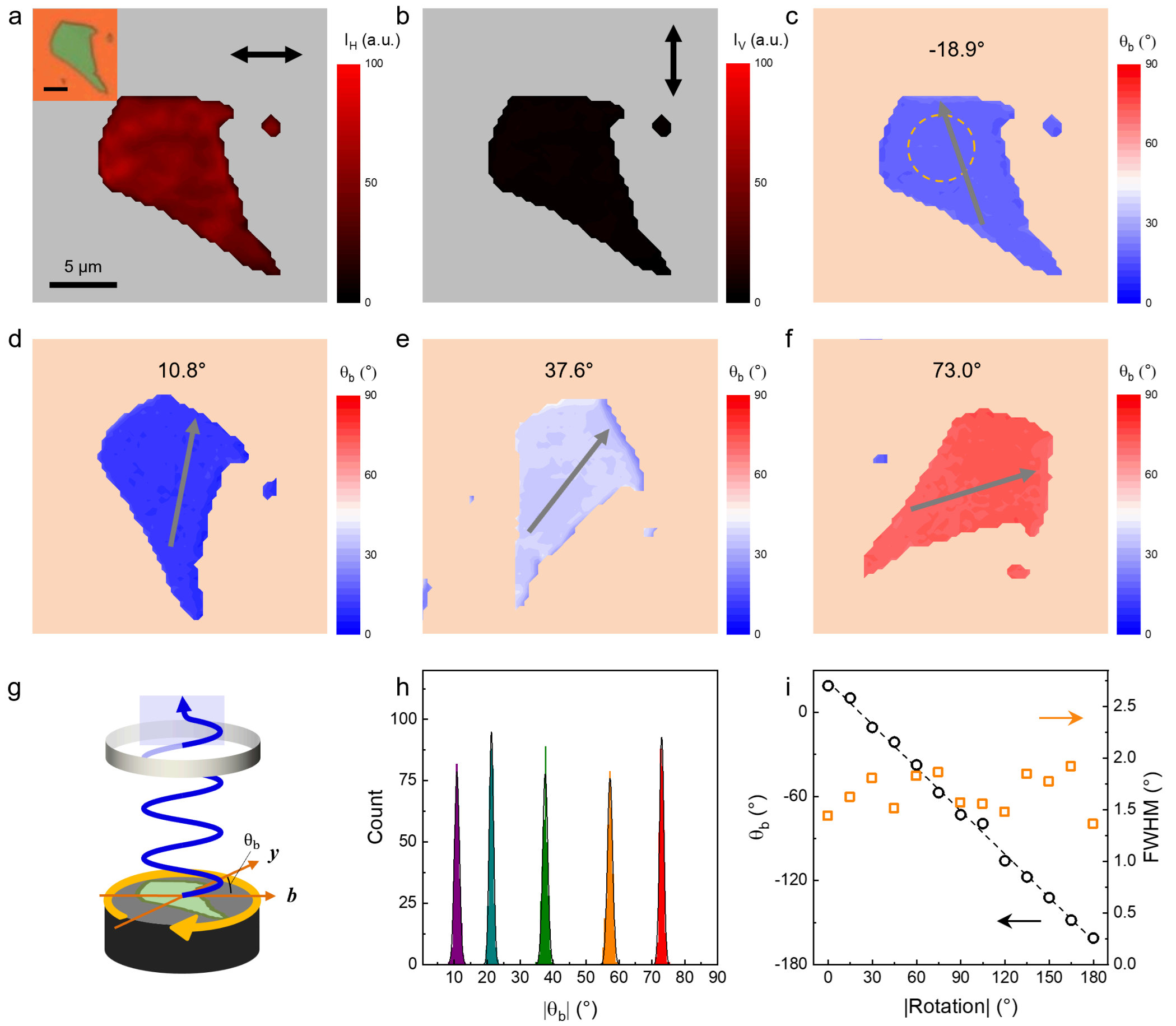

**Figure S4. Validation of crystal-orientation mapping by wide-field PL imaging.** (a,b) Polarized PL intensity ($I_H$ and $I_V$) images of an exfoliated bulk Tc crystal. The black arrows indicate the horizontal (a) and vertical (b) polarization directions of the 457 nm excitation laser. The inset in (a) shows an optical image of the ~60-nm-thick Tc crystal used for imaging. (c–f) Crystal orientation ($\theta_b$) maps measured at different sample rotation angles. Each map is derived from the respective horizontal and vertical PL intensities. Using $\theta_b$ maps obtained at different rotation angles, the ambiguity between the two mathematical solutions for $\theta_b$ was resolved. Gray arrows indicate the average crystal orientation determined from orientation histograms in (h). (g) Schematic of the wide-field PL imaging setup. We performed measurements by rotating the sample using a rotation mount (yellow arrow). (h) Histograms of the crystal orientation extracted from the middle of the sample, orange dashed circle in (c), at different rotation angles. (i) $\theta_b$ (left ordinate) and FWHM of $\theta_b$-histogram (right ordinate) as a function of rotation angle, where each histogram was well-fitted by a Gaussian function. The linearity and narrow distribution indicate the accuracy and precision of the orientational PL polarimetry.

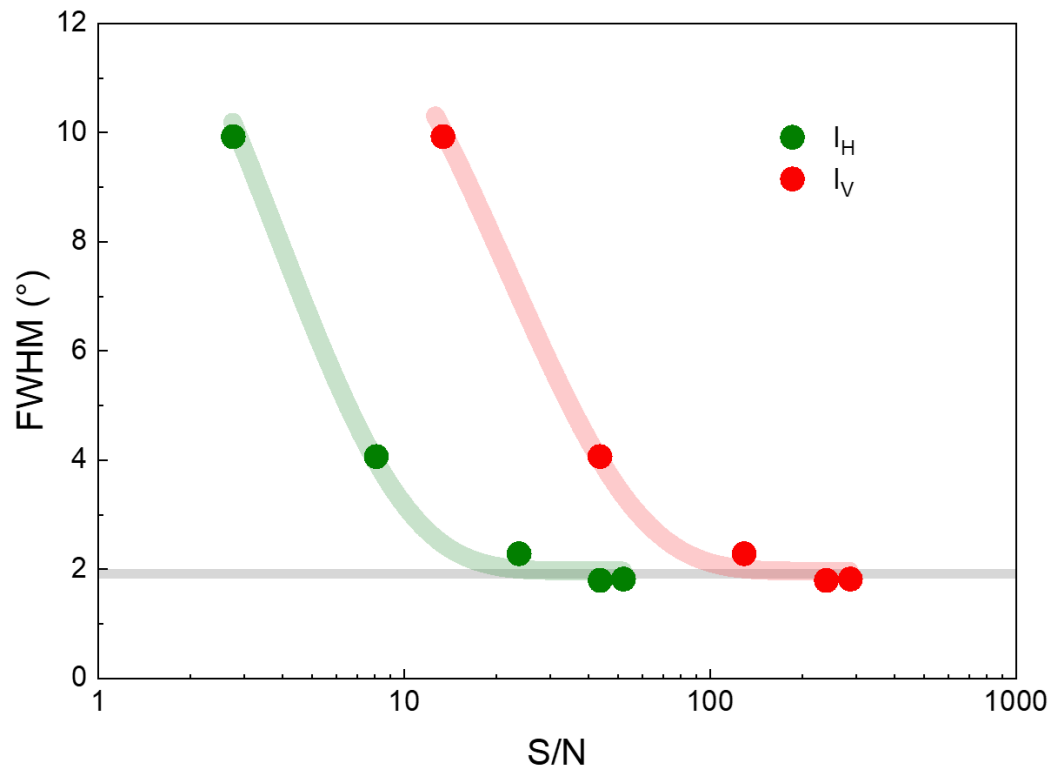

**Figure S5. Tc orientation distribution as a function of the signal-to-noise ratio (S/N) in wide-field PL imaging.** The precision of the orientation imaging was evaluated using the FWHM of orientation histograms (cf. Figs. 2f or S4h). An exfoliated 21L Tc crystal on a silicon substrate was imaged with acquisition times ranging from 1 to 100 s. The green and red circles represent the same orientation-histogram FWHM plotted against the S/N of the horizontal and vertical PL images, respectively. For both polarizations, the FWHM decreases rapidly with increasing S/N, converging to 1.9° at high S/N. For each polarization, S/N was defined as the mean PL intensity, $I_H$ or $I_V$, divided by the standard deviation of the background signal measured in a sample-free region.

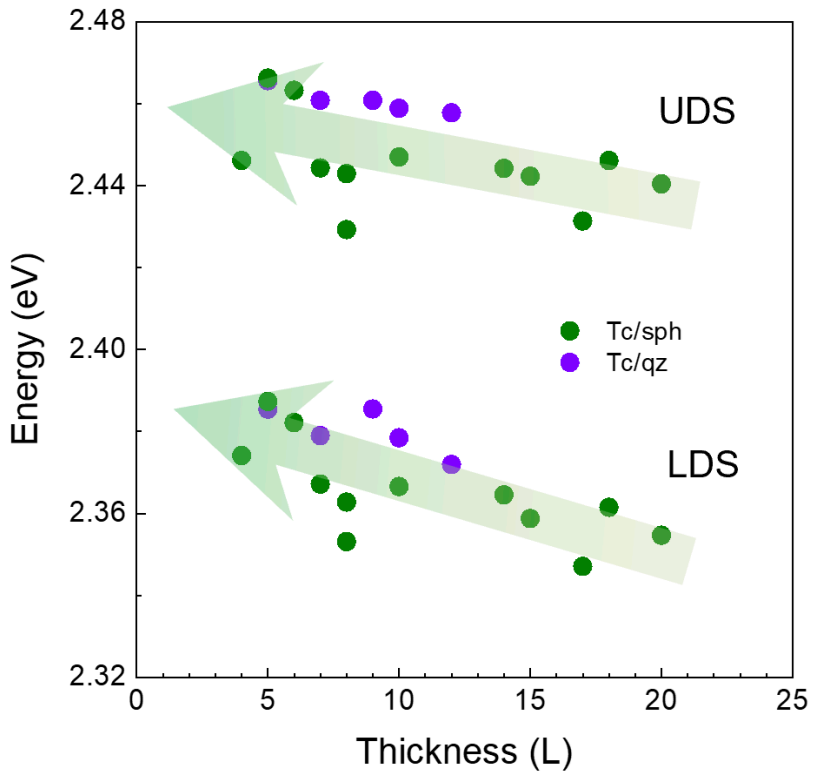

**Figure S6. Thickness dependence of the two 0-0 Davydov states in Tc.** The energies of both the LDS and UDS increase as the thickness of exfoliated Tc decreases. The LDS exhibits a larger blueshift than the UDS, reducing the DS energy toward the few-layer limit. Green arrows indicate the overall thickness-dependent trends of the LDS and UDS.

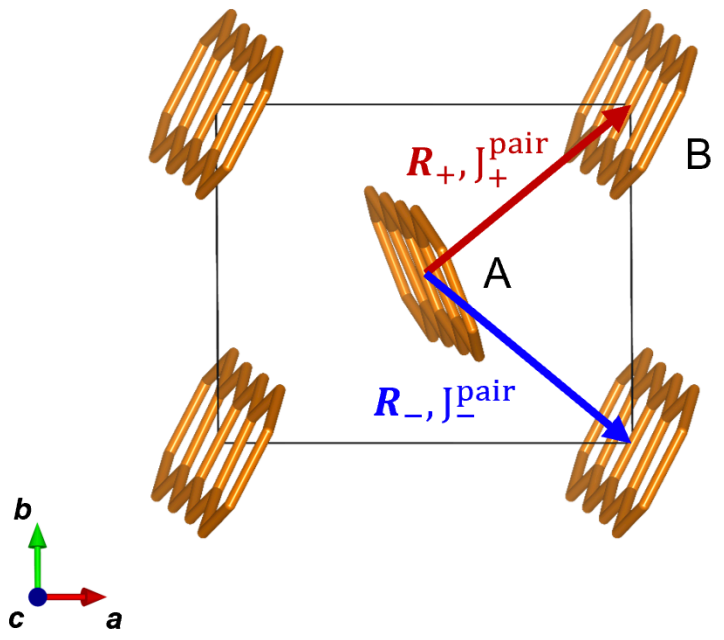

**Figure S7. Diagonal intermolecular separations used to estimate packing-dependent Coulomb couplings in Tc.** Schematic view of the molecular arrangement in the ***a***-***b*** plane. A and B denote the two molecular sublattices; the corner B molecules are shared by neighboring unit cells. From the central A molecule, the two representative center-to-center displacement vectors are $\boldsymbol{R}_{+} = (\boldsymbol{a} + \boldsymbol{b})/2$ and $\boldsymbol{R}_{-} = (\boldsymbol{a} - \boldsymbol{b})/2$, with lengths $R_{+}$ and $R_{-}$, respectively. Their pairwise Coulomb couplings are denoted by $J_{+}^{pair}$ and $J_{-}^{pair}$. The $\pm$ subscripts distinguish the two diagonal directions, not the signs of the couplings. Opposite neighbors along each diagonal have the same respective coupling in the adopted lattice model. Distance and coupling changes calculated from the mean unit-cell parameters in Table 1 of the main text are described in Eqs. S4 and S5. Transition-dipole magnitudes, orientation factors and dielectric screening are held unchanged in this estimate.

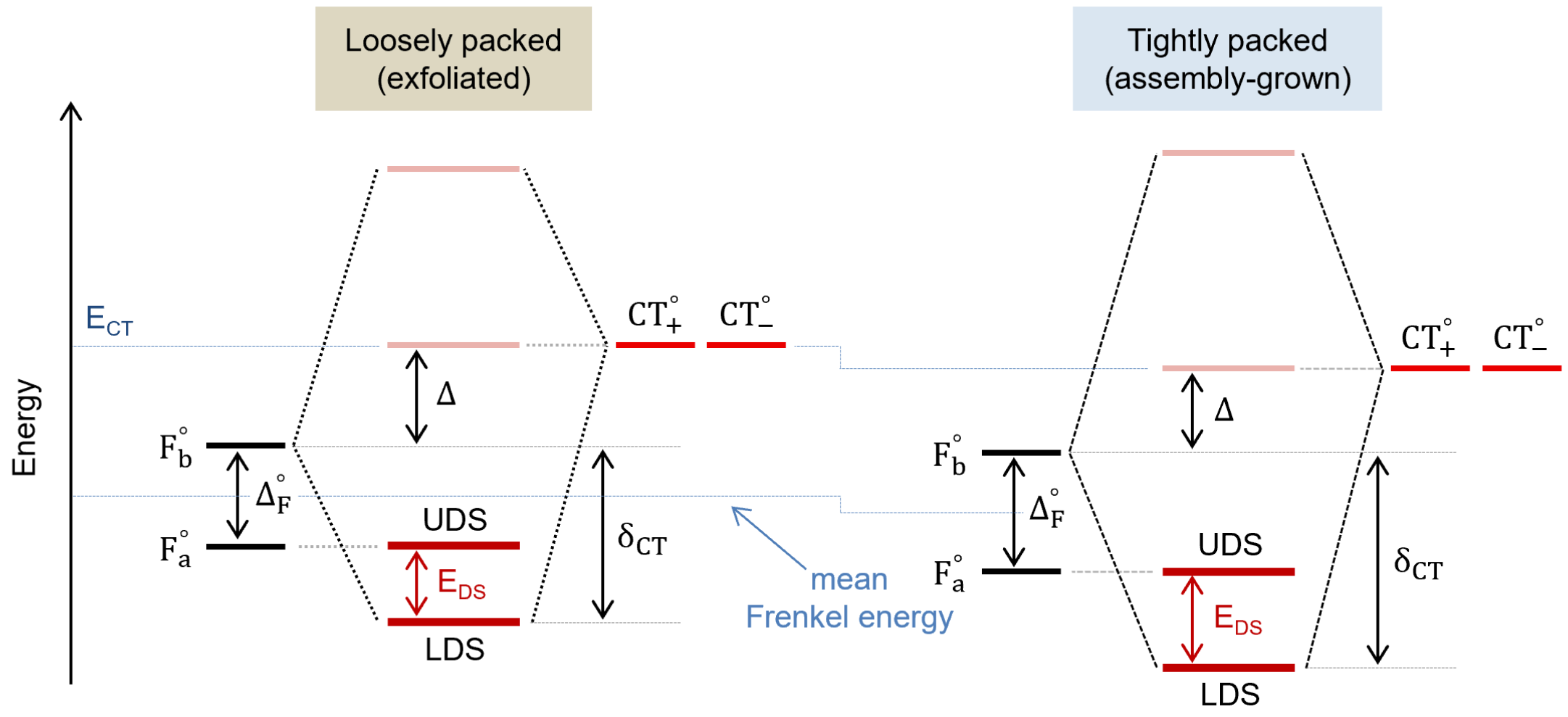

**Figure S8. Schematic interpretation of packing-dependent Davydov-state energies.** Energy-level diagrams for loosely (left) and tightly (right) packed Tc crystals, representing exfoliated and assembly-grown crystals, respectively. A lower mean Frenkel energy shifts both diabatic Frenkel states downward. Relative to this mean, an increase in the diabatic Frenkel separation $\Delta_F^o$ lowers the ***a***-polarized Frenkel state $F_a^o$, identified with the unmixed UDS, and raises the ***b***-polarized Frenkel state $F_b^o$. Lowering the diabatic CT-state energy $E_{CT}$ relative to $F_b^o$ or increasing the Frenkel-CT coupling $|V|$ can strengthen mixing and increase the CT-induced stabilization $\delta_{CT}$ of the LDS. When the increase in $\delta_{CT}$ exceeds that in $\Delta_F^o$, the LDS redshifts more than the UDS, increasing the observed DS energy $E_{DS} = E_{UDS} - E_{LDS} = \delta_{CT} - \Delta_F^o$.

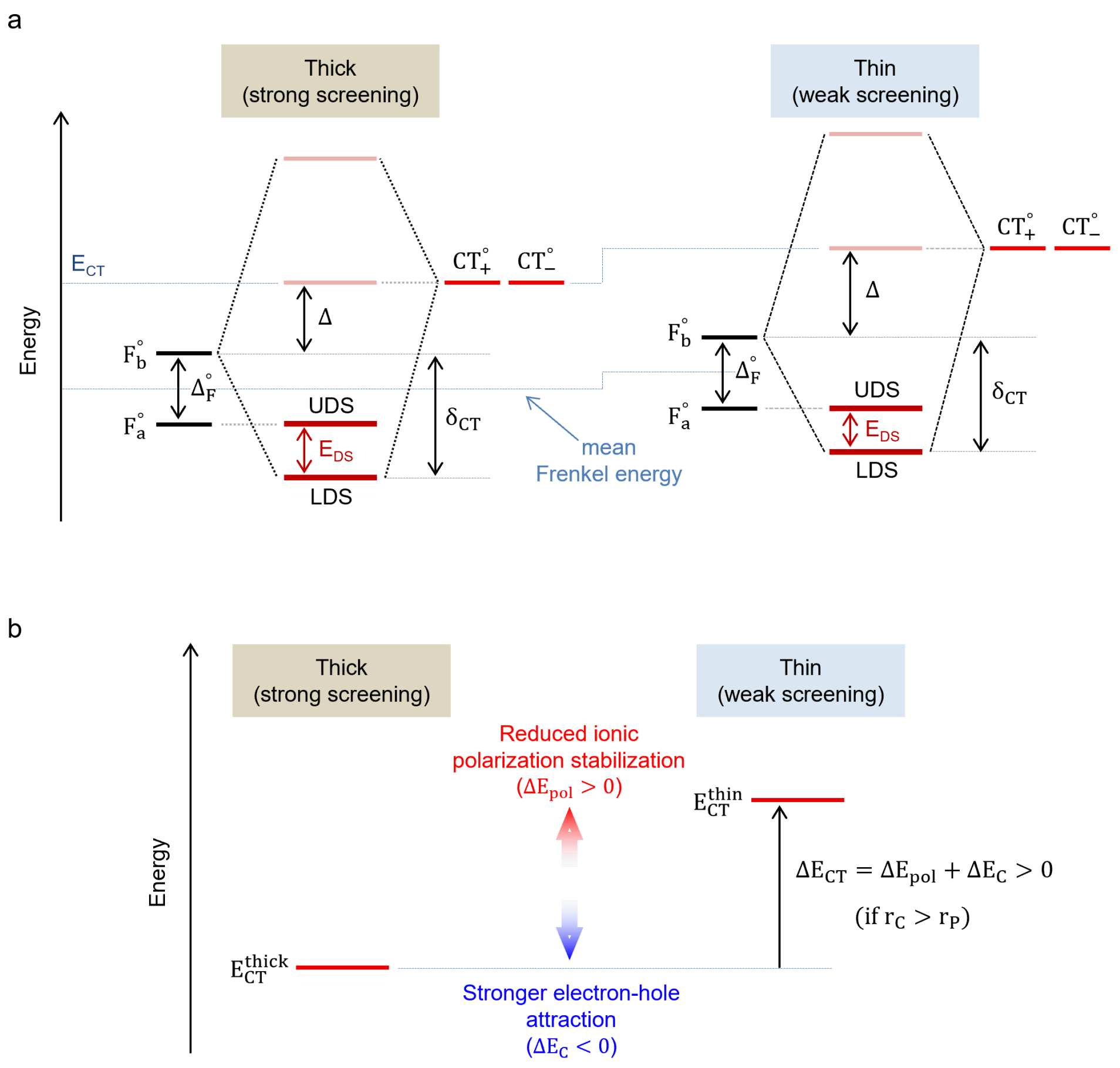

**Figure S9. Schematic interpretation of thickness-dependent Davydov-state energies.** (a) Energy-level diagrams for thick (left) and thin (right) Tc within the common-shift approximation. On thinning, the two diabatic Frenkel states shift upward by the same amount, leaving their separation $\Delta_F^o$ unchanged, while the diabatic CT state shifts upward by a larger amount. At fixed Frenkel-CT coupling $V$, the increased Frenkel-CT energy gap $\Delta = E_{CT} - E_{F,b}^o$ reduces the CT-induced stabilization $\delta_{CT}$ of the LDS. The LDS blueshifts more than the UDS, reducing the observed DS energy. The ***a***-polarized Frenkel state is treated as unmixed, and the upper CT-mixed state is shown above the diabatic CT states. (b) Competing contributions to the CT-state energy under weaker dielectric screening. Reduced ionic polarization stabilization raises the CT energy, whereas stronger electron-hole attraction lowers it. The CT energy increases when the former effect dominates, corresponding to $r_C > r_P$ in the simplified Born model (Eq. S7).

**Supplementary Note**

**Frenkel-charge-transfer (CT) mixing model**

To interpret the trends of excitonic transition energies of tetracene (Tc) crystals, we employ the well-established Frenkel-CT mixing model[1, 2] depicted by Fig. 2a of the main text. The observed energy trends and their qualitative interpretation are summarized in Table 2 of the main text; the model relations and assumptions underlying this interpretation are detailed below. The diabatic Frenkel energies are denoted by $E^{o}_{F,b}$ and $E^{o}_{F,a}$, which correspond to the predominantly ***b***- and ***a***-polarized states at k = 0 before CT mixing. Their energies include local excitation energies and Coulomb interactions across the lattice, with both positive and negative contributions.

Following the idealized monoclinic limit of Yamagata et al.[3] with vibronic coupling neglected, we take $E^{o}_{F,b} > E^{o}_{F,a}$ and equal electron and hole dissociation integrals ($D_e = D_h$). The higher Frenkel state, $F^{o}_{b}$, mixes with $CT^{o}_{+}$, the symmetric diabatic CT combination, of energy $E_{CT}$ through a CT coupling matrix element $V$. The antisymmetric Frenkel-CT coupling between $F^{o}_{a}$ and $CT^{o}_{-}$, proportional to $D_e - D_h$, vanishes, leaving $F^{o}_{a}$ unmixed. For triclinic Tc, the a/b labels specify the predominant polarization directions of the optical transitions, and do not refer to particular symmetric or antisymmetric Frenkel states.

The LDS and UDS energies are

$$E_{LDS} = \frac{E^{o}_{F,b} + E_{CT}}{2} - \frac{1}{2}\sqrt{\left(E_{CT} - E^{o}_{F,b}\right)^2 + 4|V|^2}, \qquad \text{(S1a)}$$

$$E_{UDS} = E^{o}_{F,a}. \qquad \text{(S1b)}$$

We define the Frenkel-CT energy gap as $\Delta = E_{CT} - E^{o}_{F,b} > 0$. The CT-induced stabilization of the LDS and the observed DS energy are

$$\delta_{CT} = E^{o}_{F,b} - E_{LDS} = \frac{\sqrt{\Delta^2 + 4|V|^2} - \Delta}{2}, \qquad \text{(S2a)}$$

$$E_{DS} = E_{UDS} - E_{LDS} = \delta_{CT} - \Delta^{o}_{F}, \qquad \text{(S2b)}$$

where $\Delta^{o}_{F} = E^{o}_{F,b} - E^{o}_{F,a} > 0$ is the separation between the two diabatic Frenkel states. The adopted LDS-UDS ordering corresponds to $\delta_{CT} > \Delta^{o}_{F}$.

***Packing dependence: exfoliated 2D vs grown 2D Tc.*** Compared with exfoliated Tc, assembly-grown Tc exhibits redshifts of both Davydov states, with a larger shift of the LDS and hence a larger DS energy (Figs. 3c–3e of the main text). The pairwise Coulomb coupling between the two transition dipoles in a Tc dimer, in the point-dipole approximation, is[2]

$$\mathrm{J}_{12}^{\mathrm{Coul}} = \frac{\boldsymbol{\mu}_1 \cdot \boldsymbol{\mu}_2 - 3(\boldsymbol{\mu}_1 \cdot \hat{\boldsymbol{R}})(\boldsymbol{\mu}_2 \cdot \hat{\boldsymbol{R}})}{4\pi\varepsilon_0\varepsilon'\mathrm{R}^3}, \qquad \text{(S3)}$$

where $\boldsymbol{\mu}_1$ and $\boldsymbol{\mu}_2$ are the molecular transition dipoles, $\boldsymbol{R} = \mathrm{R}\hat{\boldsymbol{R}}$ connects their centers, and $\varepsilon'$ is the effective local relative dielectric constant screening this interaction. The sign of $\mathrm{J}_{12}^{\mathrm{Coul}}$ depends on the relative dipole orientations and their alignment with $\hat{\boldsymbol{R}}$; reducing R alone increases its magnitude, making positive couplings more positive and negative couplings more negative. At fixed transition-dipole magnitudes, orientation factors and screening, the diagonal distances (Fig. S7) and fractional coupling changes are

$$\mathrm{R}_{\pm} = \frac{1}{2}\sqrt{\mathrm{a}^2 + \mathrm{b}^2 \pm 2\mathrm{ab}\cos\gamma}, \qquad \text{(S4a)}$$

$$\eta_{\pm} \equiv \frac{\mathrm{J}_{\pm}^{\mathrm{grown}} - \mathrm{J}_{\pm}^{\mathrm{exfoliated}}}{\mathrm{J}_{\pm}^{\mathrm{exfoliated}}} = \left(\frac{\mathrm{R}_{\pm}^{\mathrm{grown}}}{\mathrm{R}_{\pm}^{\mathrm{exfoliated}}}\right)^{-3} - 1, \qquad \text{(S4b)}$$

where $\mathrm{a}$, $\mathrm{b}$ and $\gamma$ are the unit cell parameters given in Table 1 of the main text. The distances $\mathrm{R}_+$ and $\mathrm{R}_-$decrease by 3.2% and 1.2%, respectively, giving $\eta_+$= 0.101 and $\eta_-$= 0.037.

In the idealized lattice model, the diabatic Frenkel separation is $\Delta_{\mathrm{F}}^{\mathrm{o}}$ = 2$\mathrm{L}_{12}(0)$, where $\mathrm{L}_{12}(0)$ is the sum of Coulomb couplings between the two molecular sublattices at k = 0, as defined by Yamagata et al.[3]. Here, the net negative contribution from Coulomb couplings within each molecular sublattice lowers both Frenkel energies equally, whereas their separation depends on the sum of inter-sublattice couplings. For comparable molecular orientations and screening, denser packing is expected to strengthen this negative contribution, lowering the mean Frenkel energy. This common shift does not directly change the diabatic Frenkel separation, but can affect the DS indirectly through the Frenkel-CT energy gap. Given the nearly bulk-like in-plane lattice parameters of exfoliated Tc (Table 1 of the main text), we use the positive nearest-neighbor couplings reported by Yamagata et al. for the corresponding diagonal inter-sublattice pairs as approximate reference values $\mathrm{J}_{\pm}^{\mathrm{pair}}$. For the diagonal pairs defined in Fig. S7, these are $J_{+}^{pair}$ = +90 cm$^{-1}$ and $J_{-}^{pair}$ = +103 cm$^{-1}$ (unscreened). The corresponding change in the nearest-neighbor contribution is estimated as

$$\delta(\Delta_{F}^{o})_{NN} \cong 4\left(\eta_{+}J_{+}^{pair} + \eta_{-}J_{-}^{pair}\right). \quad (S5)$$

The factor of four arises from the product of two factors of two: The first counts the two distinct neighbors along each diagonal, and the second follows from the relation $\Delta_{F}^{o} = 2L_{12}(0)$. This estimate gives an increase of approximately 52 cm$^{-1}$ (6.4 meV) in the nearest-neighbor contribution to $\Delta_{F}^{o}$ before dielectric screening. A net increase in $\Delta_{F}^{o}$ would reduce the DS energy at fixed $\delta_{CT}$ (Eq. S2b).

Denser packing can also lower the diabatic CT-state energies through stronger electron-hole attraction and environmental polarization stabilization[1, 4], and can enhance Frenkel-CT coupling through increased intermolecular orbital overlap. Lowering $E_{CT}$ or increasing $|V|$, with the other parameters fixed, strengthens Frenkel-CT mixing and stabilizes the LDS without shifting the unmixed UDS. Calculations for pentacene similarly show that shorter intermolecular distances can increase the DS energy through lower diabatic CT-state energies and larger electronic coupling matrix elements[5].

In this illustrative interpretation, a lower mean Frenkel energy shifts both states downward, while an increased $\Delta_{F}^{o}$ lowers the $F_{a}^{o}$ state and raises the $F_{b}^{o}$ state relative to their mean. Both contributions lower the $F_{a}^{o}$ state, redshifting the unmixed UDS. Stronger CT mixing due to a smaller Frenkel-CT energy gap and/or stronger Frenkel-CT coupling can further lower the LDS. The resulting increase in $\delta_{CT}$ can more than offset any net increase in $\Delta_{F}^{o}$, accounting for the larger LDS redshift and larger DS energy observed in grown Tc (Fig. S8; Eq. S2b).

***Thickness dependence: bulk to exfoliated 2D Tc.*** Experimentally, both the LDS and UDS blueshift upon thinning, with a larger shift of the LDS and hence a reduced DS energy (Fig. S6). The nearly bulk-like in-plane lattice parameters of exfoliated Tc (Table 1 of the main text) lead us to consider dielectric screening rather than substantial packing reconstruction. We assume that $\varepsilon'$ is less sensitive to dimensional reduction than $\varepsilon$, the effective relative dielectric constant governing CT-state stabilization.

Next, the energy of the diabatic CT states can be approximately expressed as[1, 4]

$$E_{CT} = I_P - E_A + E_C + (P_+ + P_-), \qquad \text{(S6a)}$$

$$E_C = -\frac{e^2}{4\pi r_C \varepsilon_0}\frac{1}{\varepsilon}, \qquad \text{(S6b)}$$

$$P_\pm \approx -\frac{e^2}{8\pi r_{P_\pm}\varepsilon_0}\left(1 - \frac{1}{\varepsilon}\right), \qquad \text{(S6c)}$$

where $I_P$ is the molecular ionization potential, $E_A$ is the electron affinity, $E_C$ is the electron-hole Coulomb interaction energy, and $P_+$ and $P_-$ are the ionic polarization energies. Here, $r_C$ is the distance between the cation and anion, and $r_{P_\pm}$ is the effective polarization radius in the spherical-cavity Born model. Taking $r_{P_+} = r_{P_-} = r_P$ gives

$$E_{CT} = I_P - E_A - \frac{e^2}{4\pi\varepsilon_0}\left\{\frac{1}{r_P} + \left(\frac{1}{r_C} - \frac{1}{r_P}\right)\frac{1}{\varepsilon}\right\}. \qquad \text{(S7)}$$

Weaker screening strengthens electron-hole attraction but reduces ionic polarization stabilization. At fixed molecular parameters, the latter effect dominates when $r_C > r_P$, raising $E_{CT}$ (Fig. S9b).

In the unmixed-UDS approximation, the observed UDS blueshift corresponds to an upward shift of $E^o_{F,a}$ (Eq. S1b). We adopt the common-shift approximation: The two diabatic Frenkel states are assumed to shift upward by the same amount (i.e., constant $\Delta^o_F$), and the coupling $V$ is held fixed as a simplifying assumption based on the largely preserved in-plane packing. The common-shift approximation assumes that changes in the environment shift the excitation energies of molecules in the two sublattices by similar amounts. It also neglects thickness-dependent changes in the net inter-sublattice Coulomb coupling due to altered screening or the removal of neighboring layers, thereby maintaining $\Delta^o_F$ constant. The mixed LDS responds to both $E^o_{F,b}$ and $E_{CT}$. At fixed nonzero $V$, the sensitivity of $E_{LDS}$ to $E_{CT}$ and the differential change in $E_{DS}$ are

$$w_{CT} \equiv \frac{\partial E_{LDS}}{\partial E_{CT}} = \frac{1}{2}\left(1 - \frac{\Delta}{\sqrt{\Delta^2 + 4|V|^2}}\right), \qquad \text{(S8a)}$$

$$dE_{DS} \equiv dE_{UDS} - dE_{LDS} = w_{CT}\left(dE^o_{F,b} - dE_{CT}\right) - d\Delta^o_F. \qquad \text{(S8b)}$$

Here, $0 < w_{CT} < 1/2$. The common-shift approximation sets $d\Delta_F^o = 0$ in Eq. S8b. A common upward Frenkel shift with $E_{CT}$ fixed would blueshift the UDS more than the LDS (Eq. S8b), opposite to the observed trend.

In the proposed thinning scenario, the two diabatic Frenkel energies shift upward equally, while a larger CT-state upshift under reduced screening increases the Frenkel-CT gap, weakens mixing and reduces $\delta_{CT}$. The UDS follows the common Frenkel shift, whereas the LDS blueshifts by this shift and the loss of CT-induced stabilization. At fixed $\Delta_F^o$, this additional LDS blueshift reduces $E_{DS}$ (Eq. S2b), reproducing the observed thickness dependence (Figs. S6 and S9a).

**Supplementary References**